\documentclass[pre,aps,twocolumn,groupedaddress,showpacs]{revtex4}
\usepackage{amssymb}
\usepackage{graphics}
\begin{document}
\bibliographystyle{apsrev}
\title{Pinch Resonances in a Radio Frequency Driven SQUID Ring-Resonator System}

\author{H.Prance}
\author{T.D.Clark} 
 \email{t.d.clark@sussex.ac.uk}
\author{R.Whiteman} 
\author{R.J.Prance} 
\author{M.Everitt}
\author{P.Stiffell}
\affiliation{Physical Electronics Group\\
 School of Engineering and Information Technology\\
 University of Sussex\\
 Brighton, Sussex BN1 9QT, U.K.}
\author{J.F.Ralph}
\affiliation{Department of Electrical Engineering and Electronics\\
 University of Liverpool\\
 Brownlow Hill,\\
 Liverpool L69 3GJ}

\date{30 January 2001}

\begin{abstract}
In this paper we present experimental data on the frequency domain response
of a SQUID ring (a Josephson weak link enclosed by a thick superconducting
ring) coupled to a radio frequency (rf) tank circuit resonator. We show that
with the ring weakly hysteretic the resonance lineshape of this coupled
system can display opposed fold bifurcations that appear to touch (pinch
off). We demonstrate that for appropriate circuit parameters these pinch off
lineshapes exist as solutions of the non-linear equations of motion for the
system.
\end{abstract}
\maketitle

\section{Introduction}

It is well known that the system comprising a SQUID ring (i.e. a single
Josephson weak link enclosed by a thick superconducting ring), inductively
coupled to a resonant circuit [typically a parallel LC radio frequency (rf)
tank circuit], can display very interesting dynamical behaviour~\cite{1,2,3,4,5}.
This behaviour is due to the non-linear dependence of the ring screening
supercurrent $\left( I_{\text{s}}\right) $ on the applied magnetic flux $
\left( \Phi _{\text{x}}\right) $ which originates ultimately from the cosine
(Josephson) term in the SQUID ring Hamiltonian~\cite{6}. In the past,
limitations in the noise level, bandwidth and dynamic range of rf receivers
used to investigate this coupled system have restricted the range of
non-linear phenomena that could be observed, i.e. only relatively weak
non-linearities in $I_{\text{s}}\left( \Phi _{\text{x}}\right) $ could be
probed. However, recent improvements in rf receiver design have changed this
situation quite radically.

The regimes of behaviour of SQUID rings~\cite{4} are usually characterized in
terms of the parameter $\beta =\left( 2\pi \Lambda I_{\text{c}}\right) /\Phi
_{\text{o}}$, where $I_{\text{c}}$ is the critical current of the weak link
in the ring, $\Lambda $ is the ring inductance and $\Phi _{\text{o}}\left(
=h/2e\right) $ is the superconducting flux quantum. Thus, when $\beta \leq 1$
(the inductive, dissipationless regime) $I_{\text{s}}$ is always single
valued in $\Phi _{\text{x}}$. By contrast, for $\beta >1$ (the hysteretic,
dissipative regime) this current is multi-valued in $\Phi _{\text{x}}$. Both
these regimes can lead to very strong non-linearities in $I_{\text{s}}\left(
\Phi _{\text{x}}\right)$~\cite{4,5,7}. For the inductive regime these will be
strongest when $\beta \rightarrow 1$ from below, at which point $I_{\text{s}
} $ becomes almost sawtooth in $\Phi _{\text{x}}$ (modulo $\Phi _{\text{o}}$
). For the hysteretic regime the existence of bifurcation points in $I_{
\text{s}}$, at particular values of $\Phi _{\text{x}}$, can lead to even
stronger non-linearities. In the absence of external noise these
bifurcations, linking adjacent branches in $I_{\text{s}}\left( \Phi _{\text{x
}}\right) $, occur when $\Phi _{\text{x}}=\pm \Lambda I_{\text{c}}$.

Although the behaviour of single weak link SQUID rings (as shorted turns)
can be monitored at any non-zero frequency, most studies have been carried
out at radio frequency (rf, typically $\approx 20MHz$) for which a wide
range of high performance electronics is available. We have adopted this
frequency range in the work described here. In practice, signal to noise is
improved (as in this work) by driving the SQUID ring through the
intermediary of a resonant circuit, almost invariably a parallel LC (tank)
circuit with a high quality factor $\left( Q\right) $. A schematic of the
SQUID ring, inductively coupled to an rf tank circuit (with parallel
inductance and capacitance $L_{\text{t}}$ and $C_{\text{t}}$, respectively),
is shown in figure 1. 
\begin{figure}
   \resizebox*{0.48\textwidth}{!}{\includegraphics{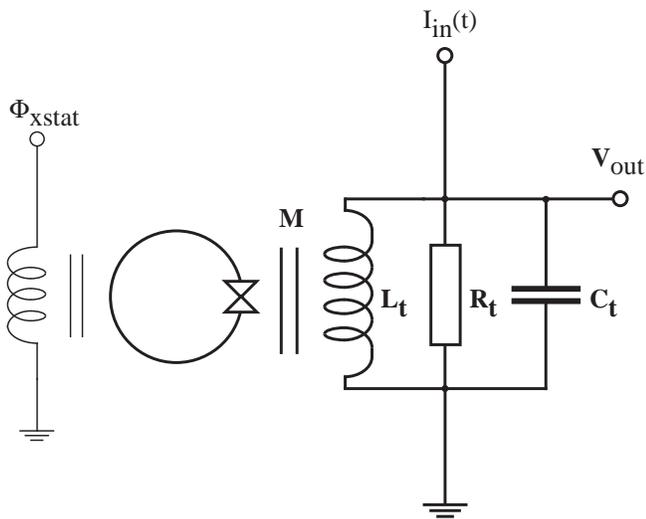}}
  \caption{Schematic of the inductively coupled SQUID ring-tank circuit
    system.}
  \label{f1}
\end{figure}
In the most commonly used experimental scheme the
dynamical behaviour of the ring is followed in the time domain through the
so-termed SQUID magnetometer characteristics. In these, the rf voltage $
\left( V_{\text{out}}\right) $ developed across the tank circuit is plotted
as a function of the rf current $\left( I_{\text{in}}\right) $ used to
excite it, where $I_{\text{in}}$ is linearly amplitude modulated in time
(figure 1). This current generates an rf flux (peak amplitude $\varphi $) in
the tank circuit inductor (coil), a fraction of which ($\mu =M/L_{\text{t}}$
for a coil-ring mutual inductance of $M$) then couples to the SQUID ring.
The supercurrent response in the ring then back-reacts on the tank circuit,
and so on. Any additional static or quasi-static magnetic flux $\left( \Phi
_{\text{xstat}}\right) $ that is required is usually supplied via a second
coil coupled to the SQUID ring, again as depicted in figure 1. For small $
\beta $-values (i.e. from $<1$ to a few) the $V_{\text{out}}$ versus $I_{
\text{in}}$ dynamics contain as their main element a set of voltage features
repeated periodically in $I_{\text{in}}$ at intervals $\varpropto \Phi _{
\text{o}}/\Lambda $. The precise form of the SQUID magnetometer
characteristics depends both on the type of rf detection employed (e.g.
phase sensitive or diode) and the value of $\Phi _{\text{xstat}}$.

The non-linear screening current response can also be made manifest in
the frequency domain. This can be accomplished at a fixed level of
(peak amplitude) rf flux $\mu \varphi $ at the SQUID ring, without the
need for amplitude modulation. For example, in the inductive regime,
with $\beta $ very close to one from below, we have demonstrated that,
at sufficiently large values of $\mu \varphi $, fold bifurcations
develop in the resonance lineshape of a SQUID ring-tank circuit
system. Indeed, using a bidirectional sweep (up and down in
frequency), and at certain values of $\mu \varphi $ and $\Phi
_{\text{xstat}}$, the system can even display opposed (hammerhead)
fold bifurcations in this lineshape, one on each side of the
resonance~\cite{7}.  Opposed bifurcations are not only most unusual
but, to our knowledge, the possibility of their existence is scarcely
mentioned in the literature on non-linear dynamics~\cite{8}. In the case
of the inductive SQUID ring, with $ \beta $ very close to one, we have
found that these bifurcations are always well separated in frequency.
However, in this work we will show that in the weakly hysteretic
regime of SQUID behaviour $\left( \beta \approx few\right) $ this
separation can become extremely, perhaps vanishingly, small.
Throughout this work we shall term this a pinch resonance.

The one sided (above or below resonance) weak fold
bifurcation~\cite{9} is taken as the text book example of the onset of
non-linear behaviour in a strongly driven oscillator. The development
of opposed fold bifurcations (on both sides of the resonance) in any
non-linear system is another matter altogether. As far as we are
aware, opposed fold bifurcations had not be seen experimentally prior
to our report of hammerhead resonances in the frequency domain of
coupled inductive SQUID\ ring-tank circuit systems~\cite{7}, and only
once in the theoretical literature~\cite{8}. In itself, we consider that
this constitutes a significant contribution to the body of knowledge
concerning non-linear dynamical systems. That the cosine non-linearity
of the SQUID ring can also generate, in part of its parameter space,
something as exotic a a pinch resonance is simply remarkable,
particularly since such resonances appear as solutions to the
established equations of motion for the ring-tank system in the weakly
hysteretic regime. What this result, and others, point to is that the
non-linear dynamics of SQUID rings coupled to other, linear, circuits
is likely to prove very rich indeed~\cite{10}, the more so since the
parameter space that has so far been explored is still very limited.
As such, these coupled systems are likely to continue to be a focus of
attention by the non-linear community, both in theory and experiment.

With the SQUID ring treated, as in this paper, quasi-classically,
there would appear to be many potential applications of its intrinsic
non-linear nature when coupled to external circuits, for example, in
logic and memory (single~\cite{11} or multi-level~\cite{12}),
stochastic resonance~\cite{13} and in safe communication systems based
on chaotic transmission/reception~\cite{14}.  However, in the last few
years there has been a burgeoning interest in exploring the properties
of SQUID rings in the quantum regime for use in quantum technologies,
particularly for the purposes of quantum computing and quantum
encryption~\cite{15,16,17,18}. This interest has been boosted by recent
experimental work on quantum superposition of states in weak link
circuits~\cite{19,20,21,22} and very strongly in the last year by reports in
the literature of experimental investigations of superposition states
in SQUID rings~\cite{23,24}. Quantum technologies will, of necessity,
involve time dependent superpositions for their operation. Where
quantum circuits such as SQUID rings are involved, this means the
interaction with external electromagnetic (em) fields or oscillator
circuit modes. Correspondingly, the effect of applied em fields
(classical or quantum) of high enough frequency and amplitude is to
generate quantum transition regions where energy is exchanged between
the ring energy levels. Given the cosine term in the SQUID ring
Hamiltonian~\cite{6}, these energy exchanges in general involve
(non-perurbatively) coherent multiphoton absorption or emission
processes~\cite{25,26}. These exchange regions extend over very small
ranges in the static magnetic flux $\Phi _{\text{xstat}}$\ applied to
the ring (typically over $10^{-3}$ to $10^{-4}\Phi _{\text{o}}$),
comparable to the width of the anticrossing regions reported in the
recent experiments on superposition of states in SQUID
rings~\cite{23,24}. Such narrow flux widths imply the concomitant
existence of extremely strong non-linearities in the screening
supercurrents flowing in the ring. In general these current
non-linearities generate non-linear dynamical behaviour in external
(classical) circuits used to probe the superposition state of the
ring. As we have emphasized in previous publications~\cite{27,28,29},
following these non-linear dynamics is one way of extracting
information concerning the quantum state of a SQUID ring, or any other
quantum circuit based of weak link devices. From our perspective,
therefore, it is of clear importance to the future development of
quantum technologies based on SQUID rings and related devices that
these non-linear dynamics be studied in great detail. Given the
results presented in this paper, the non-linear dynamics of coupled
SQUID ring-probe circuit systems can display rich and yet
unanticipated behaviour. With these rings operating in the quantum
regime this is even more likely to be the case.

\section{Experimental Pinch Resonances}

Although sometimes unappreciated, probing the non-linear dynamical behaviour
of SQUID ring-tank circuit systems makes strong demands on the high
frequency electronics used. In order to open up new non-linear phenomena in
these systems we required rf electronics which combined ultra low noise
performance, large dynamic range and high slew rate capability. In practice
we achieved this by using a liquid helium cooled (4.2 Kelvin),
GaAsFET-based, rf amplifier (noise temperature $\lessapprox $ 10 Kelvin,
gain close to 20dB) followed by a state of the art, low noise, room
temperature rf receiver. With these electronics we could choose diode or
phase sensitive detection of $V_{\text{out}}$. Since phase sensitive
detection generally provides better signal to noise, this was always our
chosen mode of operation for acquiring SQUID magnetometer characteristics.

In the work reported in this paper the SQUID rings used were of the
Zimmerman two hole type~\cite{30}. These were fabricated in niobium
with a niobium weak link of the point contact kind. This weak link was
formed by mechanical adjustment between a point and a post screw, the
latter having a plane machined end face which had been preoxidized
prior to its insertion in the SQUID block. The weak link was made in
situ at 4.2 Kelvin via a room temperature adjustment mechanism. As we
have shown in previous publications~\cite{7}, the control we can exert
using this mechanism is very good, sufficient in fact to make
reproducible point contact SQUID rings with any desired $\beta $ value
from inductive~\cite{7} to highly hysteretic~\cite{12}.

In figure 2 \begin{figure}
   \resizebox*{0.48\textwidth}{!}{\includegraphics{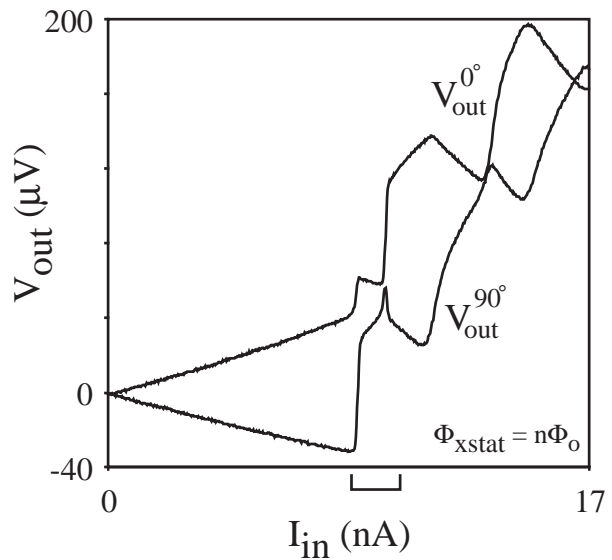}}
  \caption{Experimental equal amplitude (phase sensitive detected and $90^{
      \text{o}}$ apart in phase) rf SQUID magnetometer characteristics ($V_{out}$
    versus $I_{\text{in}}$) of a weakly hysteretic $\left( \beta =2.26\right) $
    niobium point contact SQUID ring at a static flux bias of $\Phi _{\text{xstat
        }}=n\Phi _{\text{o}}$. Here, $\Lambda =6\times 10^{-10}H$, $\omega _{\text{in
        }}/2\pi =20.122MHz$, $L_{\text{t}}=6.3\times 10^{-8}H$, $Q=1058$, $
    K^{2}=4\times 10^{-3}$ and T = 4.2 Kelvin.}
  \label{f2}
\end{figure}
we show what we term the equal amplitude magnetometer
characteristics for a weakly hysteretic ($\beta $ close to $2$) niobium
point contact SQUID ring coupled to an rf tank circuit. In these
characteristics, taken at 4.2 Kelvin with a static bias flux of $n\Phi _{
\text{o}}$, the frequency of the drive current was set to the coupled SQUID
block-tank circuit resonant frequency before a weak link contact was made
$\left( \text{i.e. }\omega _{\text{in}}/2\pi \right) $
and the voltage components ($V_{\text{out}}^{0^{\text{o}}}$ and $V_{\text{out
}}^{90^{\text{o}}}$) were maintained exactly $90^{\text{o}}$ out of phase.
However, their relative phases with respect to $I_{\text{in }}$ were rotated
electronically until the average slopes of the two characteristics are
essentially equal. This approach was adopted for convenience - it presented
the periodic features in $I_{\text{in}}$ of the orthogonal phase components
of $V_{\text{out}}$ at essentially the same amplitude. Here, the (measured)
ring inductance $\Lambda =6\times 10^{-10}H$, the measured ring-tank circuit
coupling coefficient $K^{2}\left( =\frac{M^{2}}{\Lambda L_{\text{t}}}\right)
=0.004$, $L_{\text{t}}=6.3\times 10^{-8}H$, $Q$ $=1058$\ and the drive
(excitation) frequency for $I_{\text{in}}$, $\omega _{
\text{in}}/2\pi =20.122MHz$. It is apparent from figure 2 that even
in this weakly hysteretic regime the observed SQUID magnetometer
characteristics are highly non-linear and contain a wealth of detail.

Some of the time domain information of figure 2 can be recast in a quite
dramatic form in the frequency domain. To extract this experimentally
requires the use of a high performance spectrum analyzer with a tracking
generator output~\cite{7}. This output provides a voltage signal of constant
peak to peak amplitude over the frequency range of interest. Feeding this
through a series high impedance generates an rf current in the tank circuit
which, in turn, creates an rf flux ( peak amplitude $\mu \varphi $) at the
SQUID ring. As with our previous work on SQUID ring-tank circuit dynamics in
the inductive regime~\cite{5,7}, the details of this behaviour depend crucially
on the value of the static bias flux $\Phi _{\text{xstat}}$.

In figure 3\begin{figure}
   \resizebox*{0.48\textwidth}{!}{\includegraphics{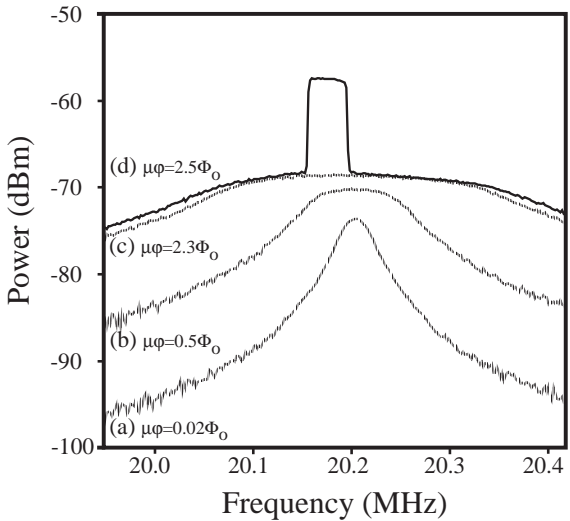}}
  \caption{(a) to (d) T = 4.2 Kelvin, $\Phi _{\text{stat}}=\left(
n+1/2\right) \Phi _{\text{o}}$ plots of the unidirectional frequency sweeps
(from low to high) for the weakly hysteretic $\left( \beta =2.26\right) $
ring-tank circuit system of figure 2 for different peak rf flux amplitudes
at the SQUID ring (i.e. $\mu \varphi =0.02,0.5,2.3$ and $2.5\Phi _{\text{o}}$
) showing the punch through at higher amplitude (e.g. $\mu \varphi
\gtrapprox 2\Phi _{\text{o}}$).}
  \label{f3}
\end{figure} we show four frequency responses for unidirectional sweeps, from
low to high frequency, for the SQUID ring-tank circuit system of figure 2
(at T = 4.2 Kelvin). The data were taken at rf flux amplitudes $\mu \varphi $
ranging from $0.02$ to $2.5\Phi _{\text{o}}$, with $\Phi _{\text{xstat}
}=\left( n+1/2\right) \Phi _{\text{o}}$. These frequency responses were
recorded using a high performance, but conventional, commercial spectrum
analyzer (a Rohde and Schwarz FSAS system) which could only scan in this
unidirectionally manner. What is clear from figure 3 is that as $\mu \varphi 
$ becomes a significant fraction of a flux quantum the lineshape of the
frequency response curves changes dramatically. In particular, for $\mu
\varphi \gtrapprox 2\Phi _{\text{o}}$ there is a punch through effect which
leads to the creation of fold bifurcation cuts. If we refer back to figure
2, this value of peak rf flux $\mu \varphi $ equates to
a drive current corresponding to the feature shown bracketed in figure 2,
but for the case of static bias flux $\Phi _{\text{xstat}}=\left(
n+1/2\right) \Phi _{\text{o}}$. The significance of this was revealed when
bidirectional frequency sweeps (from low to high and then high to low) were
taken using a spectrum analyzer of our own design. This spectrum analyzer is
shown in block form in figure 4\begin{figure}
   \resizebox*{0.48\textwidth}{!}{\includegraphics{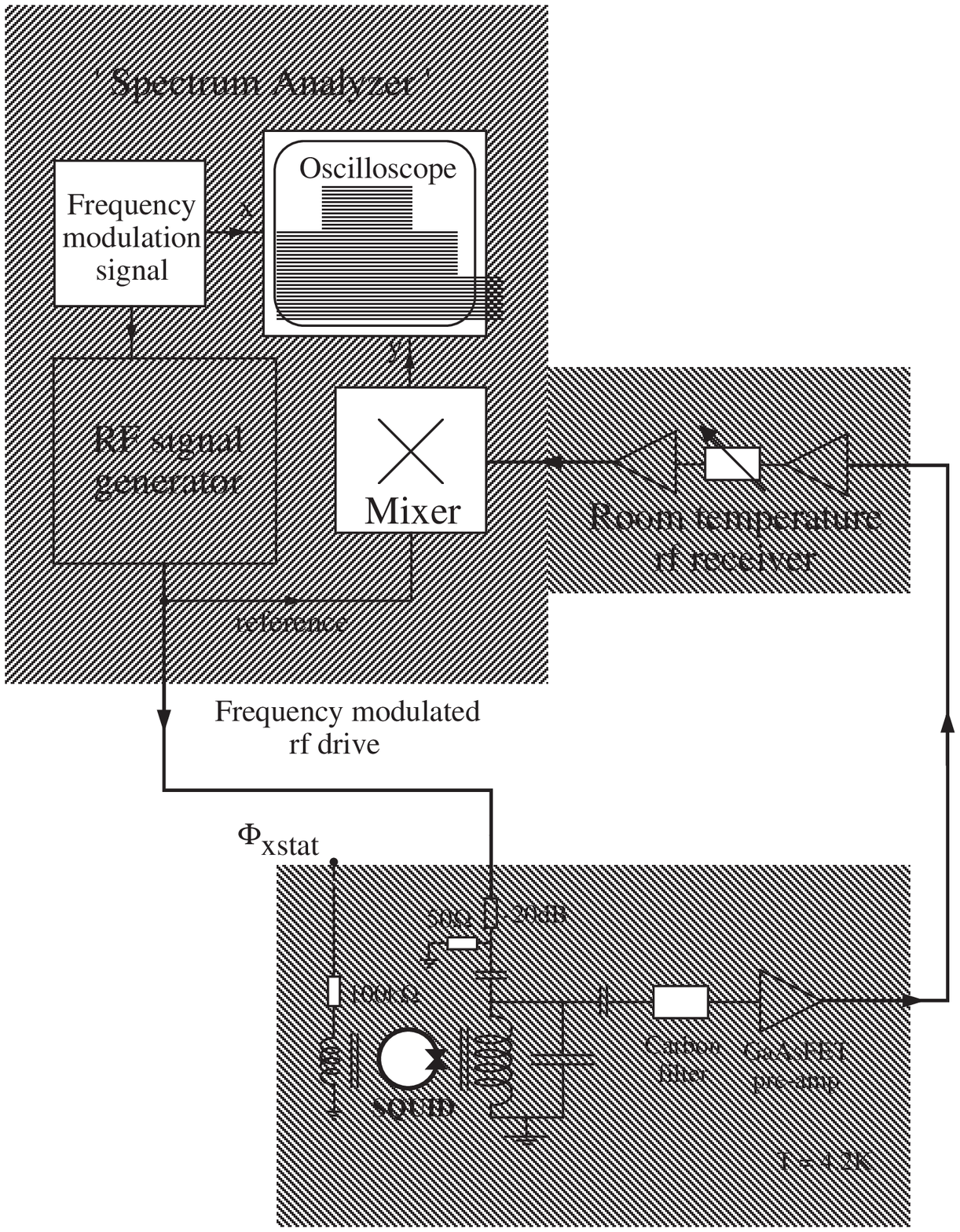}}
  \caption{Block diagram of the spectrum analyzer used to record the
bidirectional frequency response curves of figure 5. }
  \label{f4}
\end{figure}. As an illustration of its utility, we show
in figure 5\begin{figure}
   \resizebox*{0.48\textwidth}{!}{\includegraphics{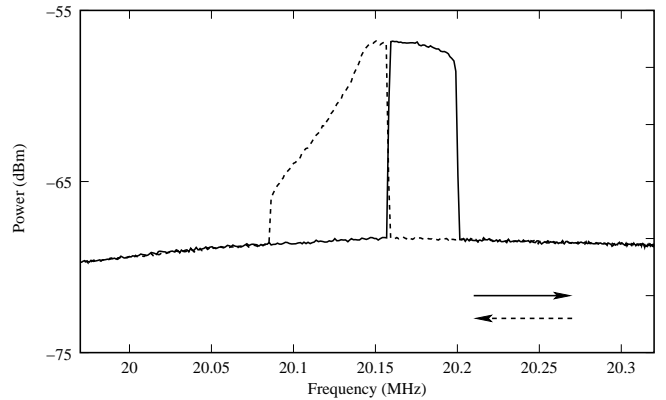}}
  \caption{Expanded, bidirectional, frequency sweeps (low to high and then
high to low) for the response (d) in the data of figure 3, again with $\mu
\varphi =2.5\Phi _{\text{o}}$ and T = 4.2 Kelvin. This value of peak $\mu
\varphi $ equates to a drive current, at static bias flux $\Phi _{\text{xstat
}}=\left( n+1/2\right) \Phi _{\text{o}}$, corresponding to the feature shown
bracketed in figure 2. }
  \label{f5}
\end{figure} the result of this bidirectional sweep for the curve
(d) of figure 3 ($\mu \varphi =2.5\Phi _{\text{o}}$), again at $
\Phi _{\text{xstat}}=\left( n+1/2\right) \Phi _{\text{o}}$. As can be seen
in figure 5, there are four fold bifurcation cuts, one each at the lowest
and highest frequency excursions in the response, and two inner cuts, one on
the sweep up and the other on the sweep down. Within the experimental
resolution, as reflected in the data points, these inner cuts appear to
overlay exactly. To our knowledge, this has not been observed before. For
this bidirectional response it seems reasonable to adopt the description
pinch resonance.

The frequency response curve of figure 5 and those plotted in
figure 3 as a function of coupled rf amplitude, are perfectly typical of the
low $\beta $ value ($\approx $ few) hysteretic SQUID ring-tank circuit
systems that we have studied. To aid in interpretation we have
provided in figure 6\begin{figure}
   \resizebox*{0.4\textwidth}{!}{\includegraphics{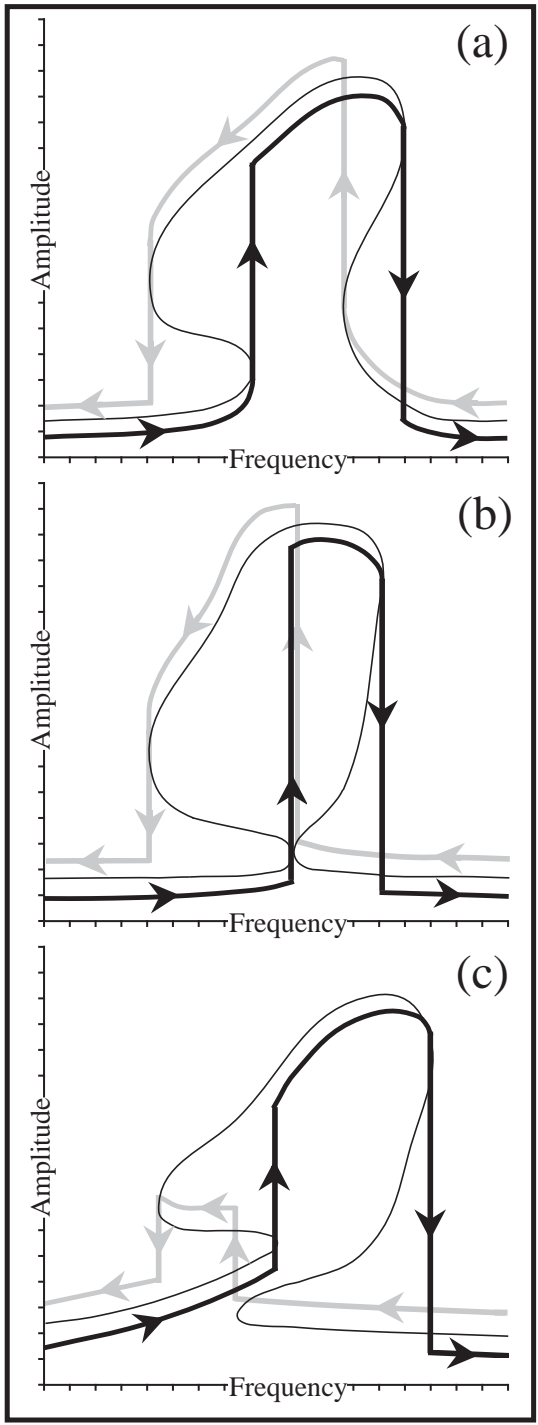}}
  \caption{A schematic of three different bidirectional frequency response
curves: (a) where the inner pair of opposed fold bifurcation cuts are well
separated in frequency (b) where these cuts superimpose in frequency (a
pinch resonance) and (c) where the fold bifurcations have moved beyond one
another. }
  \label{f6}
\end{figure} a schematic of three different bidirectional frequency
response curves: (a) where the inner pair of opposed fold bifurcation cuts
are well separated in frequency (b) where these cuts superimpose in
frequency (i.e. in our terminology, a pinch resonance) and (c) where the
fold bifurcations have moved beyond one another. It seems clear that the
data of figure 5 correspond to the bidirectional response of
figure 6(b) since this response can obviously be differentiated
from those of figures 6(a) and 6(c). It is interesting to note that the
existence of any one of the bidirectional response curves shown
schematically in figure 6 could not be revealed using standard commercial
spectrum analyzers since these always sweep the frequency unidirectionally,
i.e. from low to high. For completeness, we show in figure 7\begin{figure}
   \resizebox*{0.48\textwidth}{!}{\includegraphics{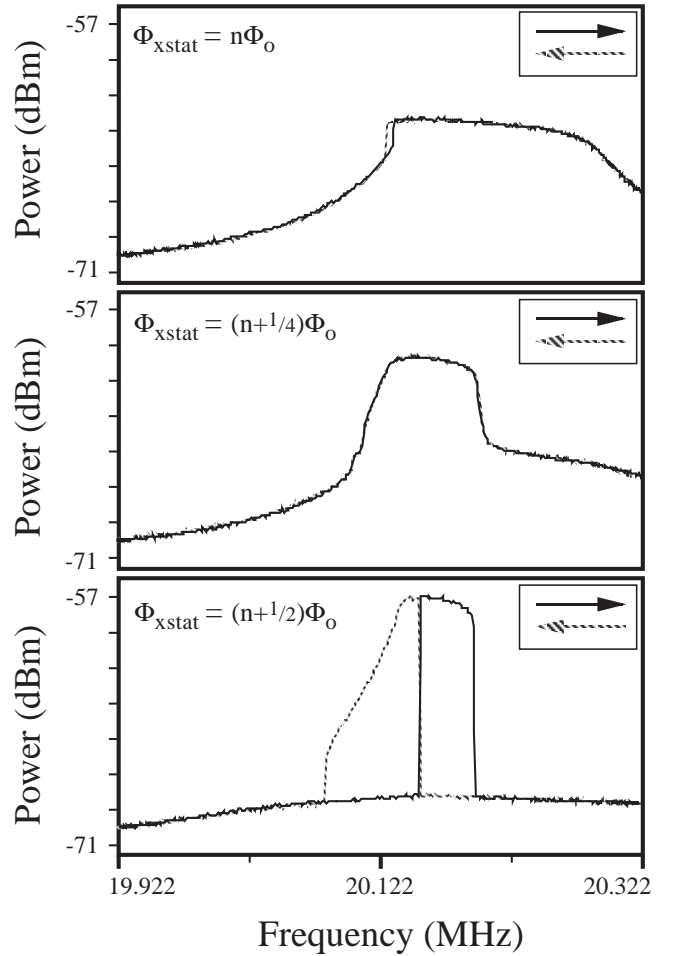}}
  \caption{Experimental bidirectional frequency response curves for the
SQUID ring-tank circuit system of figures 3(d) and 5 (T = 4.2Kelvin,
$\mu \varphi =2.5\Phi _{\text{o}}$) at three values of static bias
flux at the ring $\Phi _{\text{xstat}}=n\Phi _{\text{o}}$, $\left(
  n+1/4\right) \Phi _{\text{o}}$ and $\left( n+1/2\right) \Phi
_{\text{o}}$.}
  \label{f7}
\end{figure} the
bidirectional frequency responses (with both up and down sweeps) for the
ring-tank circuit system of figures 3(d) and 5 $\left( \text{T =
4.2 Kelvin},\mu \varphi =2.5\Phi _{\text{o}}\right) $ at the
static bias flux values $\Phi _{\text{xstat}}=n\Phi _{\text{o}}$, $\left(
n+1/4\right) \Phi _{\text{o}}$ and $\left( n+1/2\right) \Phi _{\text{o}}$.
As is apparent, the pinch resonance can only be seen in one of these
responses, i.e. at $\Phi _{\text{xstat}}=\left( n+1/2\right) \Phi _{\text{o}
} $.

\section{Theoretical Model}

\subsection{RSJ+C description}

In the standard dynamical description of a SQUID ring it is customary to
invoke the well known (quasi-classical) resistively shunted junction plus
capacitance (RSJ+C) model~\cite{4} of the weak link in the ring. Here, a pair
(Josephson) current channel is in parallel with an effective weak link
capacitance $C$ and a normal current channel of resistance $R$. For an
applied current $I>I_{\text{c}}$, the normal channel opens up with a value
of $R$ which is characteristic of the particular weak link in the ring. In
this (and other) models the SQUID ring moves in the space of total included
magnetic flux $\Phi =\Phi _{\text{x}}+\Lambda I_{\text{s}}$ with an
effective mass given by the weak link capacitance $\left( C\right) $. Within
the RSJ+C model it is always assumed that the SQUID ring can be treated as a
quasi-classical object, such that $C$ is relatively large, typically $
\approx 10^{-13}$ to $10^{-14}F$.

In the RSJ+C description the SQUID ring and tank circuit constitute a system
of coupled oscillators with equations of motion which can be written in the
form

\textit{tank circuit} 
\begin{equation}
C_{\text{t}}\ddot{\varphi}+\frac{\dot{\varphi}}{R_{\text{t}}}+\frac{\varphi 
}{L_{\text{t}}}=I_{\text{in}}+\frac{K^{2}\Phi }{M\left( 1-K^{2}\right) }
\label{tceqmot}
\end{equation}
where $R_{\text{t}}$ is the tank circuit resistance on parallel resonance of
the coupled system and $I_{\text{in}}$ is the drive current (figure 1),
which may contain both coherent and noise contributions.

\textit{SQUID ring} 
\begin{equation}
C\ddot{\Phi}+\frac{\dot{\Phi}}{R}+\frac{\Phi }{\Lambda \left( 1-K^{2}\right) 
}+I_{\text{c}}\sin \left( \frac{2\pi \Phi }{\Phi _{\text{o}}}\right) =\frac{
K^{2}\varphi }{M\left( 1-K^{2}\right) }  \label{squeqmot}
\end{equation}
where $I=I_{\text{c}}\sin \left( \frac{2\pi \Phi }{\Phi _{\text{o}}}\right) $
is the phase dependent Josephson current for the weak link in the ring.

Until recently little effort appears to have been directed to solving the
coupled equations in their full non-linear form, understandably given the
level of computational power required to deal with them. Instead it has
often been assumed that it is sufficiently accurate to linearize the SQUID
ring equation of motion. This is equivalent to invoking the adiabatic
condition that the ring stays at, or very close to, the minimum in its
potential

\begin{equation}
U\left( \Phi ,\Phi _{\text{x}}\right) =\frac{\left( \Phi -\Phi _{\text{x}
}\right) ^{2}}{2\Lambda }-\frac{I_{\text{c}}\Phi _{\text{o}}}{2\pi }\cos
\left( \frac{2\pi \Phi }{\Phi _{\text{o}}}\right)  \label{squidpot}
\end{equation}
as this varies with time. The SQUID equation (\ref{squeqmot}) then reduces to

\begin{equation}
I_{\text{c}}\sin \left( \frac{2\pi \Phi }{\Phi _{\text{o}}}\right) +\frac{
\Phi }{\Lambda \left( 1-K^{2}\right) }=\frac{K^{2}\varphi }{M\left(
1-K^{2}\right) }  \label{lineqmot}
\end{equation}

However, as we have shown in recent work~\cite{12}, this approximation should
be treated with great care, particularly in the hysteretic regime of
behaviour. In order to avoid this problem we have chosen to solve the full
coupled equations of motion [(\ref{tceqmot}) and (\ref{squeqmot})] for the
system using values of $C$ and $R$ considered typical of a weakly hysteretic
SQUID ring. We shall demonstrate that pinch resonances, such as we have
presented in figure 5, and which match with the experimentally determined
circuit parameters, show up as solutions of these equations.

\subsubsection{Pinch resonances}

With a knowledge of the system parameters, and an estimate of
$I_{\text{c}}$ , we can solve the coupled equations of motion
(\ref{tceqmot}) and (\ref{squeqmot}) numerically to find the value, at
a given $\Phi _{ \text{xstat}}$, of the rf tank circuit voltage
$\dot{\varphi}$ as a function of the frequency of the fixed amplitude
sinusoidal drive current $I_{\text{in }}$. This voltage response can
then be used to compute the bidirectional frequency responses of the
ring-tank circuit system. To do this we considered an interval $\left[
  \omega _{\text{min}}/2\pi \text{ to }\omega _{ \text{max}}/2\pi
\right] $ in the drive frequency which contained the (peak amplitude)
resonant frequency of the system. This interval was divided into $ N$
bins which gave us a resolution of $\delta \omega =\left( \omega
  _{\text{ max}}-\omega _{\text{min}}\right) /N$ in the frequency
domain. We then drove the tank circuit at $\omega _{\text{in}}$ (for
$I_{\text{in}}$)$=\omega _{ \text{min}}+n\delta \omega $ for
$n=0,1,.....,N$ increments in frequency (this being done without
changing the initial conditions of the system), thus obtaining the
power in the tank circuit at each of these frequencies via~\cite{5,7}

\begin{equation}
P\left( \omega _{\text{in}}\right) =\frac{2}{\left( 2\pi m/\omega _{\text{in}
}\right) }\left| \int_{\text{o}}^{2\pi m/\omega _{\text{in}}}\dot{\varphi}
\exp \left( i\omega _{\text{in}}t\right) dt\right| ^{2}  \label{powequ}
\end{equation}
for a given integer $m$. We note that $m$ is taken so that the power is
integrated over a few $Q$ cycles of the tank circuit oscillation to reduce
the effects of transients. In the situation where the non-linearities in the
coupled ring-tank circuit system are strong, and where, typically, the rf
drive amplitudes are large, we have already shown~\cite{5,7}that fold
bifurcation regions (i.e. regions where more than one stable solution
exists) can develop in the resonance curves of the system.

To make a comparison with the experimental data of figure 5 we estimated the
value of the $\beta $-parameter from the $0^{\text{o}}$ SQUID magnetometer
characteristic of figure 2 using the ratio (on the current axis $I_{\text{in}
}$) of the initial riser to the interval between the periodic features in
this characteristic. This yielded a $\beta $ value of $2.26$, which, for a
ring inductance of $6\times 10^{-10}H$, set $I_{\text{c}}=1.23\mu A$. In a
recent publication we described a quantum electrodynamic model~\cite{31} of a
SQUID ring which has allowed us to derive an expression for the effective
ring capacitance $\left( C\right) $, knowing the superconducting material
from which it is constructed and the value of the weak link critical
current. With niobium as the material, and $I_{\text{c}}\cong 1$ to a few $
\mu A$, this yields $C\lessapprox 10^{-14}F$. An alternative model~\cite{32},
using a different approach, gives rise to similar size capacitances. From
the literature~\cite{32,33}, a corresponding effective dissipative resistance
in the ring $\approx 100\Omega $ appears perfectly reasonable, setting a
ring $CR$ time constant of $10^{-12}$secs.

The theoretical bidirectional frequency response curves corresponding to the
experimental data (and circuit parameters) of figure 5 are shown in

figure 8\begin{figure}
   \resizebox*{0.48\textwidth}{!}{\includegraphics{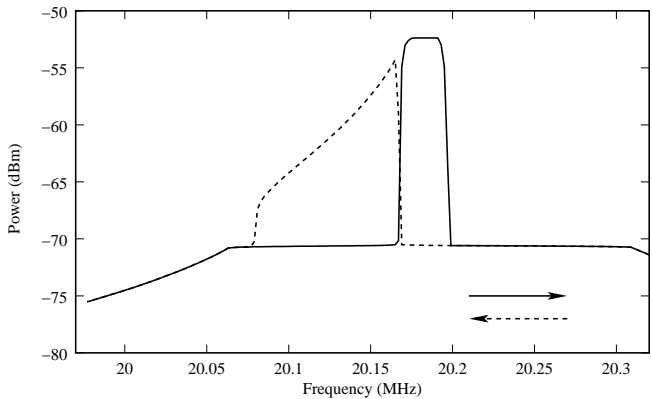}}
  \caption{Computed (RSJ+C model) bidirectional, frequency response curves
(low to high and then high to low) using the ring and tank circuit
parameters of figures 2 and 5 with $I_{\text{c}}=1.23\mu A$, $C=10^{-14}F$
and $R=100\Omega $ ($CR=10^{-12}\sec $). }
  \label{f8}
\end{figure}
, with $I_{\text{c}}$ and $CR$ taken to be $1.23\mu A$ and $10^{-12}$
secs, respectively. These were computed, sweeping forwards and backwards
across the defined frequency window, using a 4$^{\text{th}}$ order
Runge-Kutta-Merson routine with adaptive step size error control. The SQUID
ring-tank circuit coupling $\left( K^{2}\right) $ and $Q$ parameters were
set as for figures 3(d) and 5 together with $\mu \varphi =2.5\Phi _{\text{o}
} $ and $\Phi _{\text{xstat}}=\left( n+1/2\right) \Phi _{\text{o}}$. From
numerous computations we have found that the shape of the bidirectional
response in this RSJ+C approach is determined principally by the $CR$
product and is only weakly dependent on the particular values of $C$ and $R$
chosen, at least in the ranges $10^{-12}$ to $10^{-14}F$ and $1$ to $
100\Omega $. With the circuit parameters detailed above, we found that the
best fit to figure 5 (the pinch off resonance) is made with $CR=10^{-12}$
secs. As for the experimental data displayed in figure 5, the theoretical
inner fold bifurcation cuts superimpose very well, at least to the limits of
our computational accuracy. This is reflected in the computed data points in
the region of these inner cuts.

Clearly, the correspondence between the theoretically calculated response
curves of figure 8 and the experimental bidirectional responses of
figure 5 is good, both in frequency range and lineshape. For example, at $
\Phi _{\text{xstat}}=\left( n+1/2\right) \Phi _{\text{o}}$ the experimental
and theoretical responses appear to meet exactly at one point in the
frequency domain. However, we must caution that due to the numerical nature
of the calculation it is not possible theoretically to prove that this
single frequency meeting point is exact. Nevertheless, by trying different
amplitudes for $I_{\text{in}}$ we have found that we can narrow the
separation between the solutions down to an arbitrarily small level. This
reflects the situation found experimentally. We note that we have been
unable to generate this theoretical fit to our experimental bidirectional
reponse curves using the linearized SQUID equation (\ref{lineqmot}).

With regard to the alternative frequency response curves of figure 6, and
the experimental data of figures 5 and 7, we note that it may be appropriate
to treat this weakly hysteretic SQUID ring-tank circuit system as an
extremely sensitive finite state machine~\cite{34}. If so, we may be able to
compare the transition state diagram for the different cycles shown in
figure 6 [(a), (b) and (c)]. As an example, we may ask whether it is
possible in the pinch resonance (b) to go between the lower left and lower
right branches without jumping up to the loop. This could happen
occasionally but still be beyond the sensitivity limits of our present
experimental apparatus. If it does occur there will be a sharp transition
between the full pinch loop resonance [fine black line in figure 6(b)] and
the resonance displaying sets of bifurcation cuts., i.e. in the one case the
resonance loop is not accessed, in the other (the latter) it is. In
principle, the two processes could be distinguished since the pathways
around the loop involve very much larger voltages than the crossing between
the lower branches of the resonance. Such contrasting processes would, of
course, be of scientific interest but could also have potential in device
applications.

\section{Conclusions}

In this paper we have shown that a single weak link SQUID ring, weakly
hysteretic and coupled to a simple parallel resonance tank circuit, can
display quite remarkable, and previously unexpected, non-linear behaviour in
the frequency domain. Thus, we have demonstrated that this system exhibits
opposed fold bifurcation cuts in its bidirectional frequency response curves
which, at specific values of rf and static bias flux, and within our
experimental resolution, superimpose exactly. Since single fold bifurcations
in the upper or lower branches of a strongly driven resonant system are
usually taken as one of the standard examples of the onset of non-linear
dynamics~\cite{9}, the onset of opposed fold bifurcations is of special note.
That this can result, as a limiting case, in the creation of a pinch
resonance would appear to be exceptional and, in our opininion, of very
considerable interest to the non-linear community. Nevertheless, as we have
also shown, such pinch off resonances can be modelled well within
the full, non-linear RSJ+C description of the ring-tank circuit system.
Together with other recent published work~\cite{12}, these results would appear
to suggest that there is still a great deal to be explored in the
non-linear dynamics of SQUID ring-resonator systems in the quasi-classical
regime. It is also apparent that the new non-linear dynamical results
obtained in this regime point to future research needed to implement quantum
technologies based on SQUID rings (and related weak link devices)\
interacting with classical probe circuits~\cite{5,7,30}. From both perspectives
it is clear that the SQUID ring which, through the cosine term in its
Hamiltonian, is non-linear to all orders, is set to continue to reveal new
and interesting non-linear phenomena of importance in the field of
non-linear dynamics and in the general area of the quantum-classical
interface as related to the measurement problem in quantum mechanics. It is
also apparent that in both the classical and quantum regimes the non-linear
dynamics generated in coupled circuit systems involving SQUID rings could
form the basis for new electronic technologies. 

\section{Acknowledgements}

We would like to thank the Engineering and Physical Sciences Research
Council for its generous funding of this research.

\end{document}